# Binding of TMPyP$^{3+}$ porphyrin to poly(A)·poly(U) polynucleotide: a spectroscopic study

Olga A. Ryazanova, [1*] Igor M. Voloshin, [1] Igor Ya. Dubey, [2] Larysa V. Dubey, [2] Victor A. Karachevtsev [1]

[1]*Department of Molecular Biophysics, B. Verkin Institute for Low Temperature Physics and Engineering, National Academy of Sciences of Ukraine, 47 Nauky ave, 61103 Kharkiv, Ukraine.*
[2] *Department of Synthetic Bioregulators, Institute of Molecular Biology and Genetics, National Academy of Sciences of Ukraine, 150 Zabolotnogo str., 03143 Kyiv, Ukraine*
[*] Author to whom any correspondence should be addressed.

**E-mail:** ryazanova@ilt.kharkov.ua;
**Tel.:** +38-057-3410958; **Fax:** +38-057-3403370.

**ABSTRACT**

The porphyrins are macrocyclic compounds widely used as photosensitizers in anticancer photodynamic therapy. The binding of a tricationic *meso*-tris(N-methylpyridinium)-porphyrin, TMPyP$^{3+}$, to poly(A)·poly(U) polynucleotide has been studied in neutral buffered solution, pH6.9, of low and near-physiological ionic strength in a wide range of molar phosphate-to-dye ratios (*P/D*). Effective TMPyP$^{3+}$ binding to the biopolymer was established using absorption spectroscopy, polarized fluorescence, fluorimetric titration and resonance light scattering. We propose a model in which TMPyP$^{3+}$ binds to the polynucleotide in two competitive binding modes: at low *P/D* ratios (< 4) external binding of the porphyrin to polynucleotide backbone without self-stacking dominates, and at higher *P/D* (> 30) the partially stacked porphyrin *J*-dimers are embedded into the polymer groove. Enhancement of the porphyrin emission was observed upon binding in all *P/D* range, contrasting the binding of this porphyrin to poly(G)·poly(C) with significant quenching of the porphyrin fluorescence at low *P/D* ratios. This observation indicates that TMPyP$^{3+}$ can discriminate between poly(A)·poly(U) and poly(G)·poly(C) polynucleotides at low *P/D* ratios. Formation of highly scattering extended porphyrin aggregates was observed near the stoichiometric in charge binding ratio, *P/D* = 3. It was revealed that the efficiency of the porphyrin external binding and aggregation is reduced in the solution of near-physiological ionic strength.

**HIGHLIGHTS**

- Tricationic *meso*-substituted porphyrin TMPyP$^{3+}$ binds to poly(A)·poly(U) duplex in two binding modes
- The external ligand binding dominates at *P/D* < 4, whereas the embedding of the partially stacked porphyrin *J*-dimers into the biopolymer groove occurs at *P/D* > 30
- TMPyP$^{3+}$ emission enhancement is observed upon its binding to poly(A)·poly(U) in all *P/D* range
- TMPyP$^{3+}$ can discriminate between poly(A)·poly(U) and poly(G)·poly(C) polynucleotides at low *P/D* ratios
- Stable extended porphyrin aggregates are formed upon external binding

**1. INTRODUCTION**

Targeting RNA structures with small molecules and studying of molecular aspects of their complex formation is of great significance not only for fundamental science, but also for medical applications including the RNA targeted drug design [1, 2, 3, 4, 5].

The porphyrins are macrocyclic compounds with unique spectroscopic and photophysical properties owing to the presence in their structure of extended planar macrocyclic core [6, 7]. They have high biological activity conditioned in particular by their effective binding to nucleic acids (NA), where they stabilize and change the local structure [8, 9, 10]. Porphyrin – NA interaction has been the subject of multiple studies during the last four decades due to the great potential of the data obtained in biomedical, nanotechnology, and molecular electronics applications [11, 12, 13, 14, 15 ].

The water-soluble tetracationic *meso*-substituted porphyrin, TMPyP$^{4+}$ (Figure 1), is well known as a fluorescent compound that binds well to DNA and RNA of different primary and secondary structures, including single-stranded (*ss*-) [16, 17, 18, 19, 20], double-stranded (*ds*-) [8, 10, 17, 21, 22, 23], triplex [24, 25, 26] and quadruplex [27, 28, 29] ones. The biomedical significance of this porphyrin is conditioned by its ability to accumulate selectively in cancer cells, to act as an efficient antitumor agent targeting G-quadruplex structures of telomeric DNA [30, 31 32, 33, 34], and to inhibit the telomerase activity with IC$_{50}$ = 6.5 μM in the *in vitro* assay. This compound is widely used in molecular biology as a probe for the structure and dynamics of nucleic acids [8, 10, 21], as well as in medicine as a photosensitizer for photodynamic therapy of cancer [35, 36, 37], anti-viral [34, 38] and antimicrobial agent [39], as a carrier of antisense oligonucleotides for their delivery [40, 41, 42]. In complexes with redox-active transition metals (Mn$^{3+}$, Fe$^{3+}$, Cu$^{2+}$, etc.) it causes an oxidative DNA cleavage [43, 44].

Three main modes of TMPyP$^{4+}$ binding to *ds*-DNA have been established [8], which depend on the type of porphyrin substituents; availability, and the type of a metal ion coordinated in the center of the tetrapyrrole macrocycle; on the type, sequence, and structure of the target nucleic acid; on phosphate-to-dye ratio, *P/D*; on ionic strength of the solution, etc. They were identified as intercalation, groove binding, and outside binding without or with self-stacking. The last one can be accompanied by the formation of the chiral porphyrin aggregates along the biopolymer chain [16, 45] with strong interaction between the neighboring dye molecules. Since the aggregates transfer electron excitations, they can be used in material science for the creation of new photonic materials and in molecular electronics for artificial light-harvesting systems [46, 47, 48, 49]. So, it's seems appropriate to study the features of the porphyrin binding and aggregation depending on the nucleic acid composition and structural organization.

Previous studies on the interaction of TMPyP$^{4+}$ porphyrin with natural DNA and synthetic polydeoxyribonucleotides showed that this porphyrin predominantly intercalates between the GC base pairs and forms stacks on the surface of AT-rich DNA at high [porphyrin]/[DNA] ratios, but it embeds into a groove of natural or AT-rich synthetic DNA at low [porphyrin]/[DNA] ratios [50]. Investigation of TMPyP$^{4+}$ binding to different RNA duplexes and DNA·RNA hybrids [22] showed that the outside (minor groove) binding of the porphyrin with self-stacking is preferred for RNA duplexes, whereas intercalation was favored for DNA·RNA hybrids, it doesn't depend on the base composition.

*Meso*-substituted tris(*N*-methylpyridinium) porphyrin (TMPyP$^{3+}$, Figure 1) is a tricationic analog of TMPyP$^{4+}$ containing a neutral substituted phenyl group in place of one N-methylpyridinium moiety. This compound bearing a flexible and quite hydrophobic carboxyalkyl linker can be easily functionalized and covalently attached to a variety of other molecules (dyes, oligonucleotides, peptides, transport polymers, carbon nanotubes, etc.) [51, 52, 53, 54].

Earlier it was shown that this compound is also an efficient telomerase inhibitor with antiproliferative properties active at low micromolar concentrations [55]. Therefore, it is necessary to study carefully the TMPyP$^{3+}$ –NA complexes to determine the mechanism of its action and specificity. Binding of TMPyP$^{3+}$ to single-stranded inorganic polyphosphate, poly(P), [56] double-stranded poly(G)·poly(C) and four-stranded poly(G) polynucleotides [57] were studied by us using spectroscopic techniques. At low *P/D* ratios formation of extended porphyrin aggregates was detected. Binding of another tricationic derivative of TMPyP$^{4+}$ with alternating double-stranded polydeoxyribonucleotides [poly(dA-dT)]$_2$ and [poly(dG-dC)]$_2$ was studied earlier [58]

In the present work, the binding of TMPyP$^{3+}$ to poly(A)·poly(U) polynucleotide duplex has been studied in aqueous buffered solutions (5 mM phosphate buffer pH 6.9) with and without NaCl in a wide range of molar phosphate-to-dye ratios, *P/D*, using the spectroscopic techniques of absorption and polarized fluorescent spectroscopy, as well as a static light scattering. The poly(A)·poly(U) represents a right-handed A-type RNA double-helix [59]. It is known as antiviral agent and an inductor of interferon synthesis *in vivo* [60, 61, 62]. The aim of this research was to determine the types of the porphyrin binding modes to the biopolymer, and to find spectroscopic characteristics of the complexes formed depending on *P/D* ratio. Special attention is paid to features of the porphyrin aggregation near the stoihiometric binding conditions. The data obtained were compared with our previously published data [57] on TMPyP$^{3+}$ binding to *ds*-poly(G)·poly(C) and quadruplex formed by poly(G) under the same experimental conditions.

## 2. MATERIALS AND METHODS

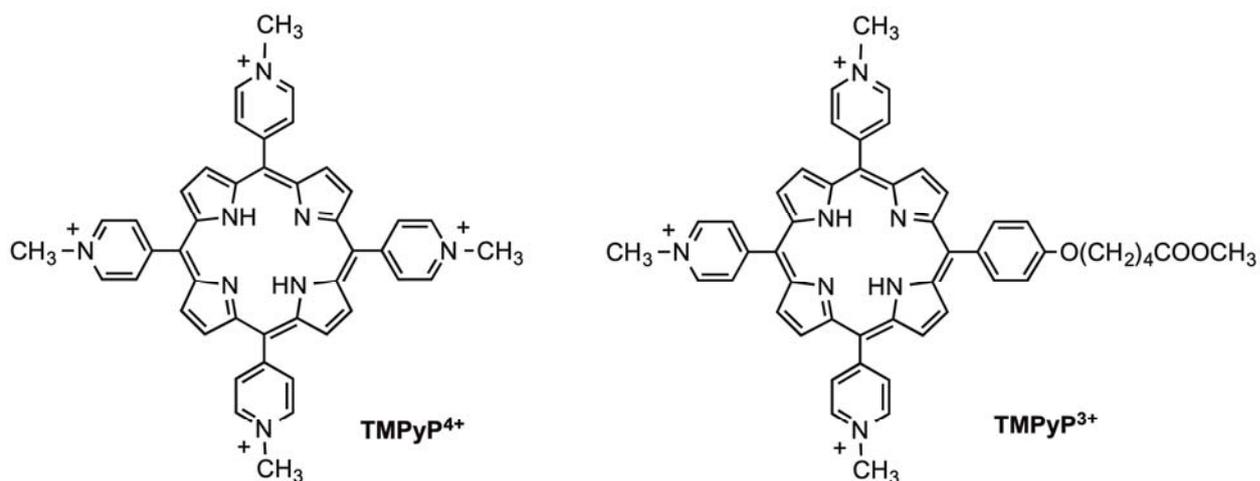

**Figure 1.** Molecular structures of tetracationic TMPyP$^{4+}$ and tricationic TMPyP$^{3+}$ porphyrins.

Tris(N-methylpyridinium-4-yl)porphyrin bearing the carboxyalkyl linker (TMPyP$^{3+}$, Figure 1) was synthesized according to the procedure described in Ref. [63]. Poly(A)·poly(U) polyribonucleotide was purchased from Sigma Chemical Co. (Europe) and used as received. The 5 mM phosphate buffer, pH 6.9, was prepared with deionized water from Millipore-Q system. To obtain the solution of near-physiological ionic strength NaCl was added to the concentration of 0.14 M. Porphyrin concentration was determined spectrophotometrically in water using the extinction coefficient of $\varepsilon_{424}$ = 200,000 M$^{-1}$ cm$^{-1}$ at the Soret band maximum [63]. Poly(A)·poly(U) concentration was determined using the extinction value of $\varepsilon_{260}$ = 7140 M$^{-1}$ cm$^{-1}$ at the ambient temperature.

Investigations were carried out using the techniques of polarized fluorescence, absorption and resonance light scattering (RLS). The experimental set-up and methods were described in detail earlier [64]. Visible electronic absorption spectra were recorded on the SPECORD M-40 spectrophotometer (Carl Zeiss, Jena). Fluorescence measurements were carried out on a laboratory spectrofluorimeter based on the DFS-12 monochromator (LOMO, 350-800 nm range, dispersion 5 Å/mm) [64]. The fluorescence was excited by the stabilized linearly polarized radiation of a halogen lamp at $\lambda_{exc}$ = 500 nm (near isosbestic point in absorption spectra). The emission was observed at a right angle to the incident beam (the spectral bandwidth was 1 nm). The fluorescence polarization degree, $p$, was calculated from the equation [65]:

$$p = \frac{F_{II} - F_{\perp}}{F_{II} + F_{\perp}} \qquad (1),$$

where $F_{II}$ and $F_{\perp}$ are intensities of the emitted light, which are polarized parallel and perpendicular to the polarization direction of the exciting light, respectively. The fluorescence spectra were corrected on the spectral sensitivity of the spectrofluorimeter.

The binding of $TMPyP^{3+}$ to poly(A)·poly(U) was followed by changes in the porphyrin fluorescence in titration experiments. Here the porphyrin sample was added with increasing amounts of the concentrated polynucleotide stock solution containing the same porphyrin content (10 μM) to achieve the desired molar phosphate-to-dye ratio, $P/D$, in the final solution. Herewith relative fluorescence intensity, $F/F_0$, and polarization degree, $p$, were measured at $\lambda_{obs}$ = 680 nm ($\lambda_{exc}$ = 500 nm).

The aggregation process arising in the solutions was followed by the intensity of the right-angle RLS excited and measured at $\lambda$ = 500 nm.

All measurements were carried out in quartz cells at ambient temperature (22 – 24°C) with freshly prepared solutions.

## 3. RESULTS

### 3.1. Absorption and fluorescence spectra of $TMPyP^{3+}$ porphyrin

The visible absorption spectrum of $TMPyP^{3+}$ presented in Figure 2 consists of the intense Soret band at 425.5 nm and four weaker Q-bands centered at 522, 561, 583.5, and 642.5 nm (Figure 2) [56]. The fluorescence spectrum of the porphyrin represents a broad featureless band (containing at least two bands) with a maximum at approximately 717 nm (Figure 3), 700 nm for non-corrected spectrum, which is typical for cationic free-base porphyrins in water [66], with a quite low value of fluorescence polarization degree ($p$ = 0.015).

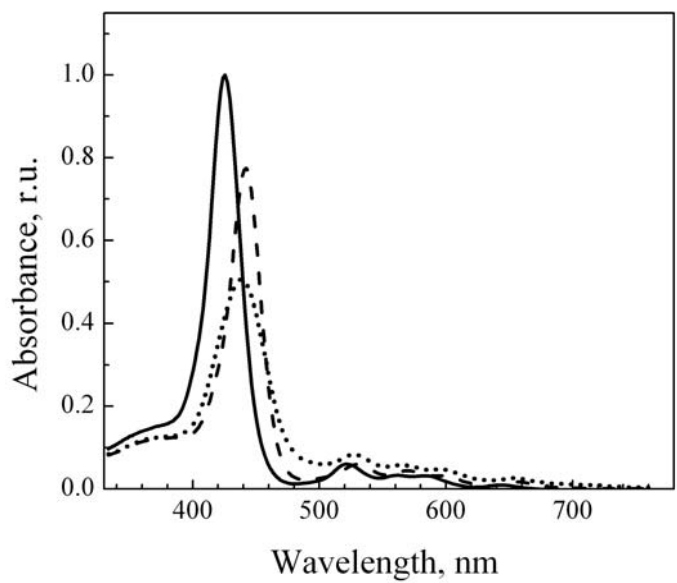

**Figure 2.** Absorption spectra of free TMPyP$^{3+}$ (solid line) and bound to poly(A)·poly(U) one at *P/D* = 3.1 (dotted line) and 1160 (dashed line). Measurements were carried out in 5 mM phosphate buffer, [C$_{dye}$] = 10 μM, and path length is 0.5 cm.

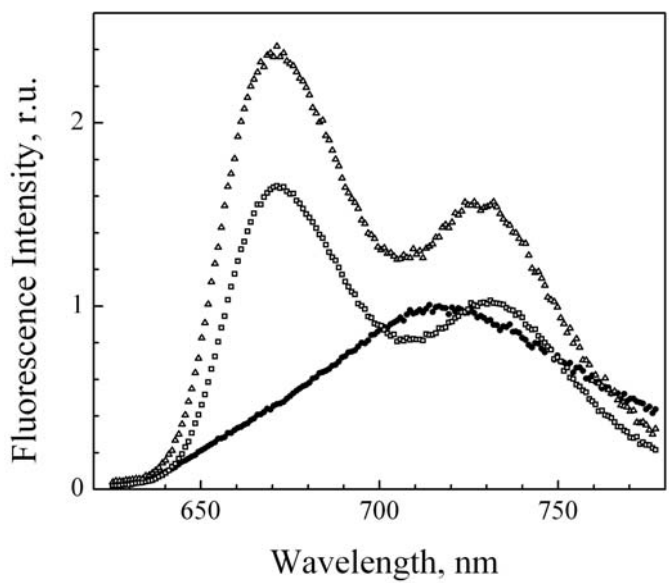

**Figure 3.** Fluorescence spectra of free TMPyP$^{3+}$ porphyrin (●) and bound to poly(A)·poly(U) one at *P/D* = 3.1 (□) and 265 (Δ). Measurements were carried out in 5 mM phosphate buffer, [C$_{dye}$] = 10 μM, λ$_{exc}$ = 500 nm.

### *3.2. Fluorimetric titration curves*

The results of fluorimetric titration of TMPyP$^{3+}$ with poly(A)·poly(U) in solutions of low and near-physiological ionic strengths (5 and ~ 148 mM Na$^{+}$, respectively) are shown in Figures 4 and 5 (the dye concentration was kept constant, 10 μM).

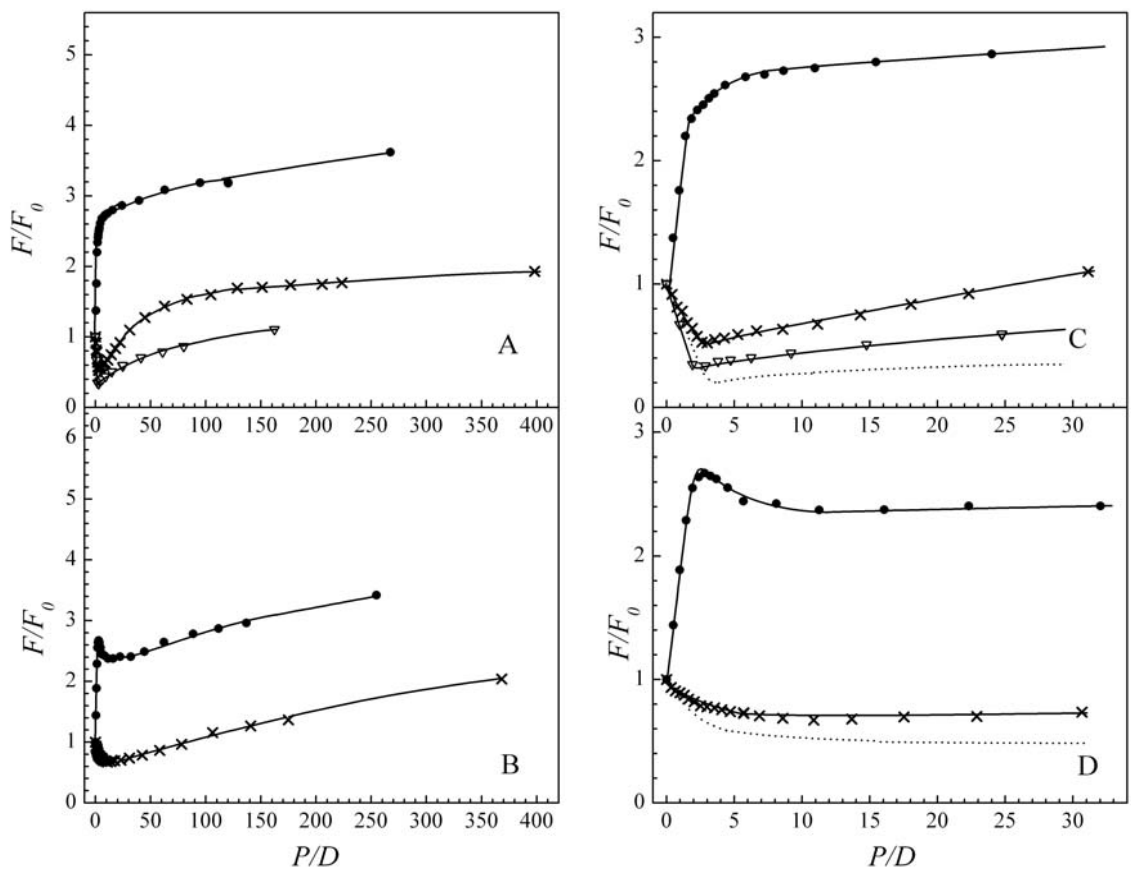

**Figure 4** Dependence of the relative intensity, $F/F_0$ ($F_0$ is the fluorescence intensity of the free dye), of TMPyP$^{3+}$ fluorescence on the molar $P/D$ ratio upon titration by poly(A)·poly(U) (●), poly(G)·poly(C) (×) [57] and poly(P) (dashed line) [64]. Measurements were carried out in 5 mM phosphate buffer (**A**,**C**), and in the buffer added with 0.148 M NaCl (**B**,**D**). The total porphyrin concentration was constant, i.e. [$C_{dye}$] = 10 μM, $\lambda_{exc}$ = 500 nm, $\lambda_{obs}$ = 680 nm. Also, the data for titration of TMPyP$^{3+}$ conjugate with imidazophenazine, TMPyP$^{3+}$-ImPzn, with poly(A)·poly(U) in the same buffer are presented (∇), [$C_{Conj}$] = 10 μM, $\lambda_{exc}$ = 440 nm, $\lambda_{obs}$ = 670 nm.

Dependence of the porphyrin relative fluorescence intensity, $F/F_0$, and polarization degree on $P/D$ ratios were registered at $\lambda_{obs}$ = 680 nm. Biphasic shapes of all titration curves presented in Figures 4(A), 4(B) indicate the existence of two different modes in porphyrin binding to the biopolymer. Note, that they are characterized by the emission enhancement. The first one dominates at a low $P/D$ ratio, whereas the second binding type prevails at $P/D > 30$.

From Figures 4(C), 4(D) presenting the initial part of titration curves in solutions of low and near-physiological ionic strengths it is seen that an increase of the biopolymer content in the range of $P/D$ = 0–2 results in the sharp 2.4–2.6-fold increase in the intensity of the porphyrin emission and sharp rise of the fluorescence polarization degree; in both cases the dependence $F/F_0$ vs $P/D$ has linear ascending character.

Then, for the sample with a low ionic strength solution (Figure 4(C)) the enhancement of the porphyrin emission slows down in the range of *P/D* = 2–5. Further increase of the polymer concentration results in only slight growth of fluorescence intensity from *F/F₀* = 2.7 at *P/D* = 7 to *F/F₀* = 3.6 at *P/D* = 267. The fluorescence polarization degree increases sharply with increasing *P/D* up to $p = 0.08$ at *P/D* = 30, then it increases slightly and polarization reaches a level of $p = 0.096$ (Figure 5).

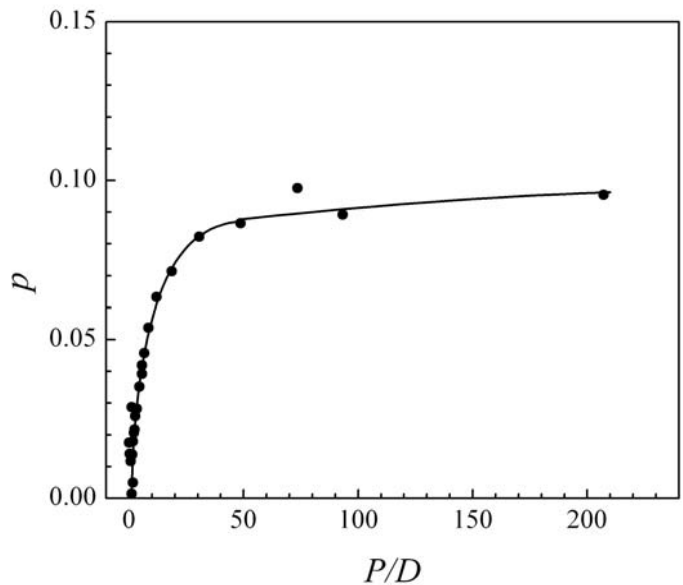

**Figure 5.** Dependence of the fluorescence polarization degree, *p*, on the molar *P/D* ratio for TMPyP$^{3+}$ bound to poly(A)·poly(U) (●). Measurements were carried out in 5 mM phosphate buffer, pH6.9, [$C_{dye}$] = 10 µM, $\lambda_{exc}$ = 500 nm, $\lambda_{obs}$ = 680 nm.

For the sample with near-physiological ionic strength solution (0.148 M Na$^+$, Figure 4(B, D)) the titration curve reaches the maximum of *F/F₀* = 2.7 at *P/D* = 2.8, then in the range of *P/D* = 2.5 - 12 the porphyrin emission is slightly weakened, to stay practically constant in the range of *P/D* = 12 – 32 at the level of *F/F₀* = 2.4, and further increase of *P/D* results in the fluorescence increase up to *F/F₀* = 3.4 at *P/D* = 255. The value of fluorescence polarization degree (not shown) reached the same steady level as it was observed in the solution containing 5 mM Na$^+$ ($p = 0.096$ at *P/D* > 100).

### 3.3. Absorption and fluorescence spectra of TMPyP$^{3+}$ bound to poly(A)·poly(U)

To clarify the spectral transformations occurring as a result of the porphyrin binding to *ds*-poly(A)·poly(U), visible absorption and fluorescence spectra of their complexes were recorded in the solution with 5 mM Na$^+$ at low and high *P/D* ratios where one of the two binding modes predominates (Figures 2 and 3). It is seen that the addition of poly(A)·poly(U) to TMPyP$^{3+}$ solution changes the shape and intensity of the porphyrin absorption and emission bands. At low *P/D* = 3.1 the Soret absorption band exhibits 49% hypochromism and 14 nm red shift (to 439 nm) (Fig. 2), whereas the fluorescence band splits into two components with maxima centered at *ca.* 671 and 728 nm, and 1.37-fold increase of its integral emission intensity occurs in comparison with that for the free dye (Figure 3).

The absorption spectrum measured at high $P/D$ = 1160 exhibits only 22% hypochromism and a more pronounced 17 nm red shift (to 442 nm) of the Soret band (Fig. 2). The maxima of the fluorescence band splitted into two components are additionally red shifted to 672 and 731 nm, and further enhancement of the porphyrin emission is observed (2.05-fold increase of the integral intensity, Figure 3). The results of our observations are summarized in the Table 1.

**Table 1.** Spectral characteristics of free TMPyP$^{3+}$ porphyrin and bound to polynucleotides one in 5 mM phosphate buffer, $\lambda_{exc}$ = 500 nm.

| Compound | P/D | Soret absorption band maxima (nm) | Normalized absorption intensity [a], $A/A_0$ | Fluorescence band maxima (nm) | Normalized fluorescence Intensity [b], $I/I_0$ | Fluorescence polarization degree [b], $p$ |
|---|---|---|---|---|---|---|
| TMPyP$^{3+}$ | 0 | 425.5 | 1 | 714 | 1 | 0.015 |
| TMPyP$^{3+}$ + poly(A)·poly(U) | 3.1 | 439 | 0.51 | 671, 728 | 2.50 | 0.025 |
| | 265 | – | – | 672, 731 | 3.62 | 0.095 |
| | 1160 | 442 | 0.78 | – | – | – |
| TMPyP$^{3+}$ + poly(G)·poly(C) [57] | 3.1 | 430 | 0.57 | 672.5, 729 | 0.52 | 0.03 |
| | 399 | – | – | 675, 731 | 1.93 | 0.08 |
| | 897 | 440.5 | 0.61 | – | – | – |
| TMPyP$^{3+}$ + poly(P) [64] | 3.5 | 443 | 0.47 | 670, 725 | 0.2 | 0.060 |
| | 115 | 426 | 0.47 | 661, 722 | 0.34 | 0.065 |

[a] in the Soret absorption band
[b] at $\lambda_{obs}$ = 680 nm

### 3.4. Resonance light scattering data

TMPyP$^{3+}$ complexes with poly(A)·poly(U) in a solution of low and moderate ionic strength exhibited strong right-angle RLS which was observed visually via a monocular tube. The dependencies of their RLS intensities on *P/D* are presented in Figure 6.

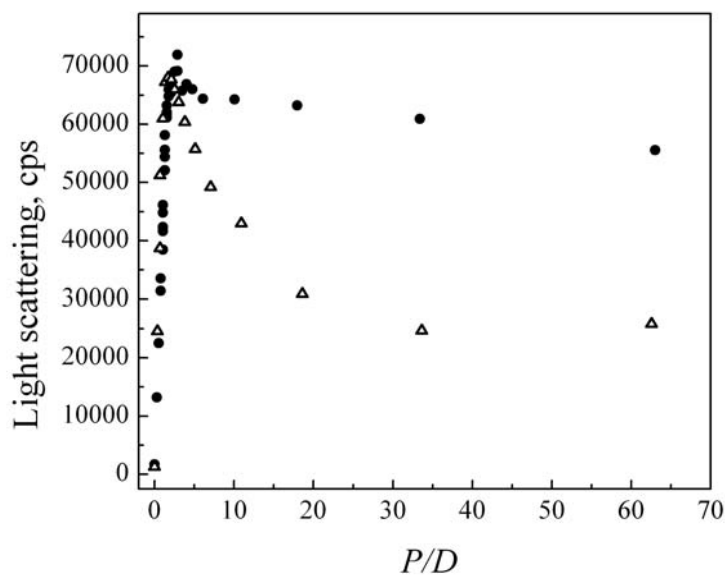

**Figure 6.** *P/D* dependence of the light scattering intensity measured in TMPyP$^{3+}$ + poly(A)·poly(U) samples at λ = 500 nm. Total porphyrin concentration was constant *i.e.* [C$_{dye}$] = 10 μM. Filled symbols correspond to the samples in 5 mM phosphate buffer without NaCl, and open ones to the same buffered solution with 0.148 M NaCl.

It is seen that in the solution of low ionic strength at low *P/D* ratios addition of the polymer induced a sharp increase in RLS intensity which reached a maximum (approximately 70,000 cps) at near-stoichiometric binding conditions (*P/D* = 3) that indicates the formation of large scale porphyrin aggregates. Further *P/D* increase results in only a slight decrease in the scattering intensity up to 55,000 cps at *P/D* = 60.

In solutions of near-physiological ionic strength the dependence of the RLS intensity *vs* *P/D* in the range of *P/D* = 0 – 3 looks like an ascending straight line which is practically coincident with that for the sample of low ionic strength. Further polymer addition results in a gradual decrease in the scattering intensity up to 25,000 cps at *P/D* = 60 due to partial aggregate disintegration. So, the largest aggregates were revealed at *P/D* = 3. From the above data it can be concluded that the aggregates formed in the solution of near-physiological ionic strength are less stable than those in the solution with low salt content.

## 4. DISCUSSION

At the initial point of the fluorimetric titration experiment (*P/D* = 0), in solution without NaCl, at a given concentration of 10 μM, the free TMPyP$^{3+}$ dye is present in monomeric form, since electrostatic repulsion between the positively charged N-methylpyridinium groups prevents formation of its dimers or multimers [67]. The emission spectrum of the free dye represents a broad featureless fluorescence band containing at least two hidden bands, which is typical for cationic free-base porphyrins in water [66] with a quite low value of the fluorescence polarization degree ($p = 0.015$).

The gradual addition of poly(A)·poly(U) to the TMPyP$^{3+}$ solution results in the substantial spectral changes which indicate that the porphyrin binds effectively to the biopolymer. The biphasic shape of fluorescence titration (Figure 4) and light scattering curves (Figure 6) indicates the existence of two binding modes: the first of them dominates at low P/D ratios (P/D < 4), whereas the second one prevails at high P/D > 30.

The spectral transformations observed at low P/D ratios are quite significant. 49% hypochromism and 14 nm red shift of the Soret absorption band (Figure 2), significant changes in the shape, position and intensity of the fluorescence band (the band splitting, 2.5-fold emission intensity increase at P/D = 4, Figures 3,4), increase in the fluorescence polarization degree up to 0.025 (Figure 5) evidences the substantial changes in the molecular environment of the porphyrin molecule. Similar transformations of absorption and fluorescence spectra were observed earlier at low P/D ratios upon binding of parent tetracationic TMPyP$^{4+}$ porphyrin to the same polynucleotide [22, 68]. Namely, 45% hypochromism and 14 nm red shift of Soret absorption band along with enhancement and splitting of the emission band into two components was reported in Ref. [68], and 44% hypochromism and 16 nm red shift of the Soret absorption band was observed in Ref. [22].

Since RNA polynucleotides, like DNA, are linear polyanions with intercharge distances practically equal to the distance between aromatic rings upon their π-π stacking, the external electrostatic binding of cationic TMPyP$^{3+}$ with or without self-stacking is supposed to be the binding type predominant at low P/D. The appearance of strong right-angle light scattering at 500 nm (so-called resonance light scattering) near the stoichiometric in charge binding ratio, P/D = 3, (Figure 6) evidences that binding to poly(A)·poly(U) stimulates the formation of extended porphyrin aggregates.

Comparison of RLS data for TMPyP$^{3+}$+poly(A)·poly(U) with those reported by us earlier for the aggregates of this porphyrin formed at low P/D ratios and under the same experimental conditions on the surface of single-stranded inorganic polyphosphate, poly(P), [56], double-stranded poly(G)·poly(C) and four-stranded poly(G) [54] exhibited similar RLS character for all these systems. It is interesting that the maximal magnitude of the RLS signal for TMPyP$^{3+}$+poly(A)·poly(U) is the same as for TMPyP$^{3+}$ + poly(P), whereas in the case of poly(G)·poly(C) and quadruplex poly(G) it is 3 and 2.5 times weaker. Upon the polymer addition, an 11% decrease of RLS intensity was observed in the P/D range from 3.5 to 25 in the case of poly(A)·poly(U) (Figure 6), and 78% for poly(G)·poly(C) [54]. In the solution with near-physiological ionic strength RLS intensity changes observed for the complexes of TMPyP$^{3+}$ with

poly(A)·poly(U) and poly(G)·poly(C) have the opposite character: increase in the relative polymer content from $P/D$ = 3 to 50 results in 63% scattering reduction in the case of the first biopolymer, and in its 4.5–fold increase for the last one.

The changes in the $TMPyP^{3+}$ fluorescence characteristics when the porphyrin binds to poly(A)·poly(U) at low $P/D$ ratios substantially differs from those observed for its complexes with *ss*-poly(P), *ds*-poly(G)·poly(C) and *fs*- (four-stranded) poly(G). For the last three polymers, strong quenching of the dye emission was observed near stoichiometric in charge $P/D$ ratios as a result of the formation of weakly fluorescing porphyrin aggregates with self-stacking of the chromophores on the polymer exterior, whereas the fluorescence of $TMPyP^{3+}$+poly(A)·poly(U) is not quenched, but on the contrary it is enhanced. The similar increase in the dye emission along with splitting of the fluorescence band into two components was observed also in the case of binding of parent $TMPyP^{4+}$ porphyrin to *ss*-poly(A) [20], poly(A)·poly(U) [68] and poly(dA)·poly(dT) [43], whereas the emission of its complexes with poly(dG)·poly(dC) was quenched [43]. It can be thus concluded that at low $P/D$ ratios, the photophysical properties of $TMPyP^{3+}$ and $TMPyP^{4+}$ bound to polynucleotides are sensitive to their base composition.

Strong resonance light scattering, enhancement of the $TMPyP^{3+}$ fluorescence upon binding to poly(A)·poly(U), sufficient hypochromism, and red shift of the Soret absorption band as well as not high magnitude of the porphyrin fluorescence polarization degree observed near stoichiometric in charge low $P/D$ ratios allow us to assume the formation of *J*-type side-by-side porphyrine aggregates without chromophore self-stacking or, less likely, the aggregates containing randomly oriented porphyrin molecules on the polynucleotide exterior.

It is interesting to note that upon the fluorescent titration of $TMPyP^{3+}$ conjugate with imidazophenazine dye ($TMPyP^{3+}$-ImPzn) with poly(A)·poly(U) under the some experimental conditions at low $P/D$ ratios we have observed a gradual decrease of the emission intensity registered for the porphyrin moiety ($\lambda_{exc}$ = 440 nm, $\lambda_{obs}$ = 670 nm.) to 34% from that for the free dye with its subsequent increase up to 110% at $P/D$ = 160, the fluorescence polarization degree increases up to 0.1 (unpublished data). Also, fluorescence quenching was observed upon binding of metallated $ZnTMPyP^{4+}$ porphyrin with axial water ligands to poly(A)·poly(U) [69].

Under the polymer excess, at $P/D$ > 30, we have observed the reduction of the hypochromism of the Soret absorption band compared to that at low $P/D$ from 49 to 22%, and further increase in the red shift of this band to $\Delta\lambda$= 17 nm, that indicates considerable perturbation of the porphyrin π-electronic system upon binding due to changes in the environment of bound chromophores. Also, more than 3.5–fold increase of the dye fluorescence at $\lambda_{obs}$ = 680 nm as compared with that for the free dye was observed along with further rise of the fluorescence polarization degree reaching the steady level of $p \leq 0.1$. The fluorescence band remains splitted into two components. The wavelengths of emission band maxima for TMPyP$^{3+}$ complexes with poly(A)·poly(U) and poly(G)·poly(C) at high *P/D* ratios practically coincide with each other and only 1-3 nm are red shifted relatively to those positions at low *P/D* ratios. And emission band maxima for TMPyP$^{3+}$ + poly(A)·poly(U) are slightly red shifted relatively to them for TMPyP$^{4+}$ + poly(rA) ($\Delta\lambda$ = 9 and 6 nm correspondingly).

Similar splitting and enhancement of the porphyrin emission band at high *P/D* ratios was observed for TMPyP$^{4+}$ bound to poly(dA)·poly(dT) [43], although the magnitudes of the corresponding red shift of the Soret absorption band was substantially lower ($\Delta\lambda$ = 8 nm). The positions of the fluorescence band maxima reported for TMPyP$^{4+}$ bound to poly(dA)·poly(dT) at high *P/D* ratios are substantially blue shifted ($\Delta\lambda$ = 14-15 nm) in comparison with those for poly(dG)·poly(dC) [43]. At the same time, the fluorescence band maxima for TMPyP$^{3+}$ complex with poly(A)·poly(U) at high *P/D* are only slightly blue shifted ($\Delta\lambda$ = 2-5 nm) in comparison with those for poly(G)·poly(C) [57].

As was established earlier [8, 9], at high *P/D* ratios the possible types of tetracatopnic TMPyP$^{4+}$ porphyrin binding to DNA are intercalation between nucleic bases and embedding into the groove of the biopolymer helix (mostly in the minor groove), and the magnitudes of their apparent binding constants are of the same order. The way of binding depends on the steric factors, type of the nucleic acid, nucleic base composition, etc. For example, a porphyrin bearing bulky peripheral substituents or containing axial ligands at the central metal ion of porphyrin metalocomplexes cannot intercalate and incorporate into the DNA groove (mainly in the minor groove). Also, it is known that TMPyP$^{4+}$ intercalates between the GC base pairs, whereas it binds at the groove of the polymer with AT sequences. One of the indicators in the determination of the binding mode is the behavior of the porphyrin Soret absorption band upon the polymer addition, because it is very sensitive to change in the local environment of the macrocycle. The empirical criteria for that were summarized in Ref. [9]. So, the intercalative binding type is characterized by a large hypochromism (H > 30%) and substantial bathochromic shift ($\Delta\lambda$ > 15 nm) of the Soret absorption band, whereas for the groove binding the spectral transformations are much less pronounced. However, these criteria were established for the dye bound to B-form DNA and polydeoxyribonucleotides. The poly(A)·poly(U) polyribonucleotide studied in the present work forms A-type double helix and thus differ by another spatial structure and characteristics [59]. Its minor groove is shallow and wide while the major groove is narrow and deep as compared to DNA grooves, and the adjacent phosphate groups in one strand locate closer to each other [59]. The grooves in DNA and RNA contain the same chemical groups, except the sugar 2'-H in DNA is replaced by 2'-OH in RNA and the thymine 5-$CH_3$ group is replaced by H in uracil [59]. In RNA duplexes the base pairs are displaced considerably from the helical axis, thereby base planes are tilted away from the plane perpendicular to the axis. These features should prevent the porphyrin chromophores from incorporation between the planes of nucleobases of RNA. In this way, intercalation of a large porphyrin molecule should be unfavored. Note, that the minor groove in the RNA duplex is wider and shallower than that in the DNA duplex; the adjacent phosphate groups in one strand are closer [59], and the groove is wide enough to accommodate the self-stacked porphyrin molecules.

For the identification of the porphyrin binding type a very sensitive tool of polarized fluorescence spectroscopy was used. Since the rotational motion of the intercalated porphyrin ligand is significantly restricted, the binding is usually accompanied by a substantial rise in its fluorescence polarization degree [65]. Despite quite high values of hypochromism and bathochromic shift of TMPyP$^{3+}$ Soret absorption band observed upon its binding to poly(A)·poly(U) at high *P/D* ratios (22 % and 17 nm correspondingly), the magnitude of fluorescence polarization degree ($p < 0.1$) remains not very high. For example, for cationic and neutral derivatives of another porphyrin, Pheophorbide-*a*, intercalated into *fs*-poly(G) at similar experimental conditions *p* reached the level of 0.29 and 0.26 respectively [70, 71]. For highly symmetric metalloporphyrins Pd(II)TMPyP$^{4+}$ [72] and Cu(II)TMPyP$^{4+}$ [73] intercalated into calf thymus DNA the rise of *p* up to ~ 0.2 were described for the first compound and more than 0.3 for the second one. Although it is known, that the symmetry of metallated porphyrin macrocycle ($D_{4h}$) is higher than that for unmetallated ones ($D_{2h}$), and the limiting theoretical value of fluorescence polarization degree for such compounds is less than 1/2, it was calculated as equal to 1/7 [7,75,76]. Thereby not high value of fluorescence polarization degree allows us to suggest that at high *P/D* ratios the molecules of tricationic porphyrin bearing the carboxyalkyl chain embed into the groove of polynucleotide duplex as *J*-dimers, like in the case of poly(G)·poly(C) [57]. This hypothesis is in a good agreement with data reported in [74] where the dependence of binding modes on the structure of uncharged *meso*-substituents was studied for several tricationic analogs of TMPyP$^{4+}$ containing a substituted phenyl group instead of N-methylpyridinium residue. Their binding to calf thymus DNA was investigated in aqueous buffered solution, pH 7.2, containing 50 mM NaCl using absorption and fluorescence spectroscopy, fluorescent titration, circular dichroism, thermal denaturation techniques and viscosimetry, along with quantum chemical calculations. One of compounds (#4 in Ref [74]) containing a relatively short alkylcarboxy side chain was quite similar to TMPyP$^{3+}$. For this derivative, the intercalative binding to DNA at high *P/D* values was not found, but availability of the groove binding was clearly established. At the same time, intercalative binding mode was determined for the porphyrin bearing the shortest side chain (OH group in the phenyl ring, compound #2 in Ref [74]), although the Soret band red shift value for this compound was determined as only 10 nm. In the same work it was found that apparent binding constants for tricationic porphyrins are two orders of magnitude lower than that for the parent tetracationic porphyrin TMPyP$^{4+}$, and their values are decreased with the extension of the side group length.

The addition of NaCl into the studied solution up to a near-physiological value of 0.148 M Na$^+$ practically doesn't change the shape and position of bands in absorption and fluorescence spectra of both free and bound porphyrin, but slightly increases their intensity for TMPyP$^{3+}$ complexes with poly(A)·poly(U). So, an increase in the solution ionic strength somewhat weakens the porphyrin electrostatic binding to nucleic acids due to the competitive binding of Na$^+$ ions to the polymer phosphate backbone, and reduces the stability of the porphyrin aggregates at high *P/D* ratios.

It is interesting to note that for another porphyrin, monocationic derivative of Pheophorbide-*a* (CatPheo-*a*), emission band doesn't split upon its binding to poly(A)·poly(U) [70], significant quenching of the dye emission up to $F/F_0 = 0.09$ is observed at low *P/D* ratios, and it remains quenched at high *P/D* ($F/F_0 = 0.2$) where fluorescence maximum is 15 nm red shifted as compared to that for the free dye. The values of fluorescence polarization degree, *p*, were 0.095 at low and 0.21 at high *P/D* ratios, which allowed us to conclude the intercalation of CatPheo-*a* molecules into the poly(A)·poly(U) double helix. So, binding of the porphyrins CatPheo-*a* and TMPyP$^{3+}$ to poly(A)·poly(U) occurs in a completely different way.

## 5. CONCLUSIONS

The present experimental study demonstrates that in a neutral aqueous solution of low and near-physiological ionic strength water-soluble tricationic *meso*-substituted porphyrin TMPyP$^{3+}$ binds effectively to double-stranded poly(A)·poly(U) polyribonucleotide via two binding modes. First of them, the external electrostatic binding without chromophores self-stacking, dominates at *P/D* < 4 ratios. The second one ascribed to the embedding of partially stacked porphyrin *J*-dimers into the groove of the polyribonucleotide helix predominates at *P/D* > 30. Absorption hypochromism and large red shift of the Soret absorption band, enhancement of the dye fluorescence, as well as by increased value of the fluorescence polarization degree were observed for both binding types. Near stoichiometric in charge binding ratio *P/D* = 3 an appearance of strong resonance light scattering was detected which indicates the formation of extended porphyrin aggregates on poly(A)·poly(U) surface. The dependence of RLS intensity on *P/D* shows that the aggregates are quite stable and do not disintegrate even at high *P/D* ratios. It was concluded that these aggregates belong to *J*-type side-by-side ones without self-stacking or contain randomly oriented porphyrin molecules. An increase in NaCl content in the solution up to near-physiological ionic strength reduces the efficiency of porphyrin external binding and the stability of aggregates at high *P/D* ratios.

From the analysis of data obtained it was established that the photophysical properties of TMPyP$^{3+}$ are sensitive to the polynucleotide base composition and can discriminate between poly(A)·poly(U) and poly(G)·poly(C) polynucleotides at low *P/D* ratios. For the first biopolymer the binding results in emission enhancement, whereas for the second one the fluorescence is quenched.

Our findings provide new insight into the features of molecular interactions between macrocyclic porphyrin dyes and nucleic acids. The results can be used in the design of porphyrin-based ligands with predictable affinity and specificity.

**ACKNOWLEDGMENTS**

Works on the study of the binding of tricationic porphyrin to nucleic acids was initiated by Dr. V.N. Zozulya, who, unfortunately, passed away.

Authors acknowledge the National Academy of Sciences of Ukraine for the financial support (Grants No. 0123U100628 and 0123U100998). O.R. acknowledges Wolfgang Pauli Institute (WPI), Vienna, Austria for the financial support (Pauli Postdoc research training scholarship in the field of Data analysis in molecular biophysics in the context of the WPI thematic programs "Quantitative Methods in Biology and Medicine" (2021/22) and "Numerical models in Biology and Medicine" (2023/2024). Also O.R. acknowledges *Nanophotonics* journal, De Gruyter, Sciencewise Publishing, and the Optica Foundation for the financial support (Ukraine Optics and Photonics Researcher Grant 2023).

**COMPLIANCE WITH ETHICAL STANDARDS**

The authors declare that our manuscript complies with all Ethical Rules applicable for this journal and that there are no conflicts of interests.

**ORCID IDs:**

Olga Ryazanova: https://orcid.org/0000-0002-8277-8611

Igor Voloshin: https://orcid.org/0009-0004-1898-2230

Igor Dubey: https://orcid.org/0000-0003-4023-4293

Larysa Dubey: https://orcid.org/0000-0001-9010-8696

Victor Karachevtsev: https://orcid.org/0000-0003-4580-6465